\documentclass[prb,superscriptaddress,twocolumn]{revtex4-1}
\usepackage[utf8]{inputenc}
\usepackage{amsmath}
\usepackage{amsfonts}
\usepackage{amssymb}
\usepackage{graphicx}
\usepackage{hyperref}
\usepackage{xcolor}
\usepackage{multirow}

\begin{document}

\title{Magnetic structures and quadratic magnetoelectric effect in LiNiPO$_4$ beyond 30$\,$T}
\date{\today}

\author{Ellen Fogh}
\thanks{ellen.fogh@epfl.ch}
\affiliation{Department of Physics, Technical University of Denmark, DK-2800 Kongens Lyngby, Denmark}
\affiliation{Laboratory for Quantum Magnetism, Institute of Physics, École Polytechnique Fédérale de Lausanne, 1015 Lausanne, Switzerland}
\author{Takumi Kihara}
\affiliation{Institute for Materials Research, Tohoku University, Sendai 980-8577, Japan}
\author{Rasmus Toft-Petersen}
\affiliation{Department of Physics, Technical University of Denmark, DK-2800 Kongens Lyngby, Denmark}
\author{Maciej Bartkowiak}
\affiliation{Helmholtz Zentrum Berlin f{\"u}r Materialien und Energie, D-14109 Berlin, Germany}
\author{Yasuo Narumi}
\affiliation{Institute for Materials Research, Tohoku University, Sendai 980-8577, Japan}
\author{Oleksandr Prokhnenko}
\affiliation{Helmholtz Zentrum Berlin f{\"u}r Materialien und Energie, D-14109 Berlin, Germany}
\author{Atsushi Miyake}
\affiliation{The Institute for Solid State Physics, University of Tokyo, Kashiwa, Chiba 277-8581, Japan}
\author{Masashi Tokunaga}
\affiliation{The Institute for Solid State Physics, University of Tokyo, Kashiwa, Chiba 277-8581, Japan}
\author{Kenichi Oikawa}
\affiliation{Materials and Life Science Division, J-PARC Center, Japan Atomic Energy Agency, Tokai, Ibaraki 319-1195, Japan}
\author{Michael Korning Sørensen}
\affiliation{Department of Physics, Technical University of Denmark, DK-2800 Kongens Lyngby, Denmark}
\author{Julia Cathrine Dyrnum}
\affiliation{Department of Physics, Technical University of Denmark, DK-2800 Kongens Lyngby, Denmark}
\author{Hans Grimmer}
\affiliation{Research with Neutrons and Muons, Paul Scherrer Institut, 5232 Villigen PSI, Switzerland}
\author{Hiroyuki Nojiri}
\affiliation{Institute for Materials Research, Tohoku University, Sendai 980-8577, Japan}
\author{Niels Bech Christensen}
\affiliation{Department of Physics, Technical University of Denmark, DK-2800 Kongens Lyngby, Denmark}


\begin{abstract}

Neutron diffraction with static and pulsed magnetic fields is used to directly probe the magnetic structures in LiNiPO$_4$ up to $25\,\mathrm{T}$ and $42\,\mathrm{T}$, respectively. By combining these results with magnetometry and electric polarization measurements under pulsed fields, the magnetic and magnetoelectric phases are investigated up to $56\,\mathrm{T}$ applied along the easy $c$-axis. In addition to the already known transitions at lower fields, three new ones are reported at $37.6$, $39.4$ and $54\,\mathrm{T}$. Ordering vectors are identified with ${\bf Q}_{\mathrm{VI}} = (0,\frac{1}{3},0)$ in the interval $37.6-39.4\,\mathrm{T}$ and ${\bf Q}_{\mathrm{VII}} = (0,0,0)$ in the interval $39.4-54\,\mathrm{T}$. A quadratic magnetoelectric effect is discovered in the ${\bf Q}_{\mathrm{VII}} = (0,0,0)$ phase and the field-dependence of the induced electric polarization is described using a simple mean-field model. The observed magnetic structure and magnetoelectric tensor elements point to a change in the lattice symmetry in this phase. We speculate on the possible physical mechanism responsible for the magnetoelectric effect in LiNiPO$_4$.

\end{abstract}

\maketitle

\section{Introduction}

The fields of study centered on multiferroics and magnetoelectrics span both fundamental physics and application with their potential for low-energy dispersive data storage and other multifunctional devices \cite{eerenstein2006,cheong2007,rivera2009,fusil2014}. In materials displaying a magnetoelectric (ME) effect, an external electric or magnetic field can induce a finite magnetization or electric polarization, respectively. The effect is usually described using Landau theory where the electric polarization, $P_i$, induced by an applied magnetic field, $H_j$, is written as \cite{eerenstein2006}
\[
P_i = P_0 + \alpha_{ij} H_j + \frac{1}{2} \beta_{ijk} H_j H_k + ...,
\]
where $i,j,k \in \lbrace a,b,c \rbrace$ and $P_0$ is a spontaneous polarization. In a similar way, the induced magnetization, $M_i$, may be expressed as follows: 
\[
M_i = M_0 + \alpha_{ji} E_j + \frac{1}{2} \gamma_{ijk} E_j E_k + ...,
\]
where $E_j$ is now the applied electric field and $M_0$ is a spontaneous magnetization. The linear ME coupling is described by $\alpha_{ij}$ and the coefficients, $\beta_{ijk}$ and $\gamma_{ijk}$, account for the quadratic ME effect. Higher order terms may also occur. The allowed ME tensor forms are governed by the magnetic symmetry of the system and $\beta_{ijk}$ has the same symmetry as the pyroelectric tensor.

In multiferroics with a strong coupling between magnetic and electric order, the mechanism is often explained by spin currents \cite{katsura2005}, the inverse Dzyaloshinskii-Moriya interaction \cite{sergienko2006} or $p$-$d$ hybridization \cite{kim2014} -- the former two are rooted in non-collinear magnetic order breaking spacial inversion symmetry \cite{mostovoy2006,kimura2007}. Examples are incommensurate spiral magnets such as the rare earth manganites $R$MnO$_3$ ($R=$ Gd,Tb,Dy) \cite{kimura2003,goto2004,kenzelmann2005} and $R$Mn$_2$O$_5$ ($R=$ Tb,Ho,Dy) \cite{hur2004,blake2005} or copper based compounds such as LiCu$_2$O$_2$ \cite{park2007} or LiCuVO$_4$ \cite{schrettle2008}. However less common, some magnetoelectric (ME) materials have magnetic order where the magnetic unit cell coincides with the crystallographic unit cell. Among these are e.g. tetragonal Ba$_2$CoGe$_2$O$_7$ \cite{kim2014} and Cr$_2$O$_3$ \cite{brockhouse1953,astrov1961}. Another example is the lithium orthophosphates, Li$M$PO$_4$ with $M=$ Ni, Co, Mn, Fe. These orthorhombic compounds (space group \textit{Pnma}) all have commensurate antiferromagnetic ground states below their respective ordering temperatures \cite{mays1963,santoro1966,santoro1967}. Although the magnetic orders have similar symmetry, the spin orientation differs depending on the magnetic ion in question due to the single-ion anisotropy. For instance, in LiNiPO$_4$ the spins are along $c$ and in LiFePO$_4$ they are along $b$ . The variations in spin orientation result in different ME tensor forms. For LiNiPO$_4$ the elements $\alpha_{ac}, \alpha_{ca} \neq 0$ are finite whereas for LiFePO$_4$ the elements $\alpha_{ab}, \alpha_{ba} \neq 0$ are finite \cite{mercier}. Previously, the field and temperature dependencies of the field-induced electric polarization in LiNiPO$_4$ \cite{jensen2009_2,toftpetersen2017} and LiFePO$_4$ \cite{toftpetersen2015} have been succesfully described based on related models.

\begin{figure}[t!]
	\centering
	\includegraphics[width = 0.9\columnwidth]{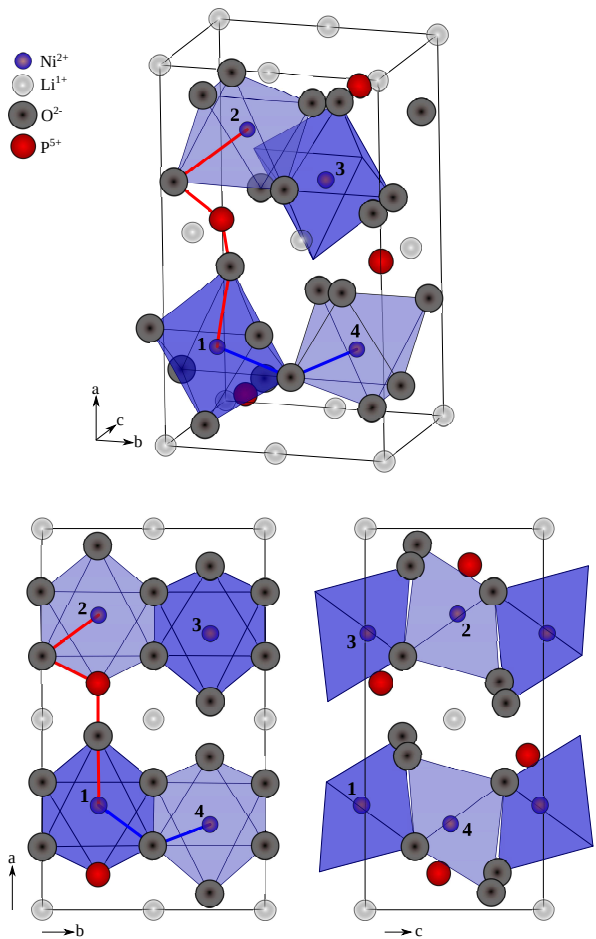}
	\caption{Crystal structure of LiNiPO$_4$. The magnetic Ni$^{2+}$ ions are surrounded by oxygen ions in an octahedral environment. The unit cell contains 4 magnetic ions which form buckled sheets in the $(b,c)$-plane. The ion positions are ${\bf r}_1 = (1/4+\varepsilon, 1/4, 1-\delta)$, ${\bf r}_2 = (3/4+\varepsilon, 1/4, 1/2+\delta)$, ${\bf r}_3 = (3/4-\varepsilon, 3/4, \delta)$ and ${\bf r}_4 = (1/4-\varepsilon, 3/4, 1/2-\delta)$ where $\varepsilon = 0.0256$ and $\delta = 0.0175$ \cite{abrahams1993}. The exchange interactions are mediated via couplings such as Ni-O-Ni (blue path) and Ni-O-P-O-Ni (red path).}
	\label{fig:structure}
\end{figure}

In this paper we focus on LiNiPO$_4$ which displays a cornucopia of magnetic phases. The crystallographic unit cell contains 4 magnetic Ni$^{2+}$ ions ($S=1$) placed in a nearly face-centered arrangement [see Fig. \ref{fig:structure} with positions \cite{abrahams1993} given in the figure caption]. Below $T_N = 20.8\,\mathrm{K}$ the spins order in an antiferromagnetic commensurate structure with propagation vector ${\bf Q}_{\mathrm{I}} = (0,0,0)$. The major spin component is along $c$ and with symmetry $(\uparrow \uparrow \downarrow \downarrow)$ \cite{santoro1966}. Here $\uparrow$/$\downarrow$ denotes spin up/down for ions on sites 1-4 following the enumeration of Ref. \onlinecite{jensen2009_2}. A smaller spin canting component along $a$ with symmetry $(\uparrow \downarrow \downarrow \uparrow)$ was also reported \cite{jensen2009_2}. Just above $T_N$ an incommensurate, linearly modulated phase exists in the narrow temperature interval up to $21.7\,\mathrm{K}$ \cite{vaknin2004, jensen2009_2, toftpetersen2011}. Upon applying a magnetic field along the easy $c$-axis, the material goes through a series of magnetic phase transitions: at $12\,\mathrm{T}$ it enters an incommensurate spiral phase with spins in the $(a,c)$-plane and propagating along $b$ \cite{jensen2009_2}. At $16\,\mathrm{T}$ the spiral locks in to a period of 5 crystallographic unit cells. Upon further increasing the field, at $19.1\,\mathrm{T}$ the spiral gives way to another ${\bf Q}_{\mathrm{IV}} = (0,0,0)$ structure which yet again at $20.9\,\mathrm{T}$ yields to a longer-period structure with a modulation of 3 unit cells along $b$ \cite{toftpetersen2017}. The magnetization in this phase is $\sim \frac{1}{3}$ of the saturated value. THz absorption spectra recorded up to $33\,\mathrm{T}$ along $c$ show changes in the magnon absorption that coincides with the magnetic phase boundaries \cite{peedu2019}. Phases I and IV (in field intervals $0-12\,\mathrm{T}$ and $19.1-20.9\,\mathrm{T}$) both support the ME effect which has previously been characterized and successfully modelled \cite{mercier,jensen2009_2,khrustalyov2016_Ni,toftpetersen2017}.

The highest field at which the magnetic structures in LiNiPO$_4$ has hitherto been probed by neutron diffraction is $30\,\mathrm{T}$ \cite{toftpetersen2017}. In this paper we combine magnetometry, electric polarization measurements and neutron diffraction to investigate the magnetic and ME phases in LiNiPO$_4$ up to $56\,\mathrm{T}$. Three magnetic phases are discovered in addition to those already known -- one of which displays a quadratic ME effect.

\section{Experimental details}

\begin{figure*}
	\centering
	\includegraphics[width = 0.8\textwidth]{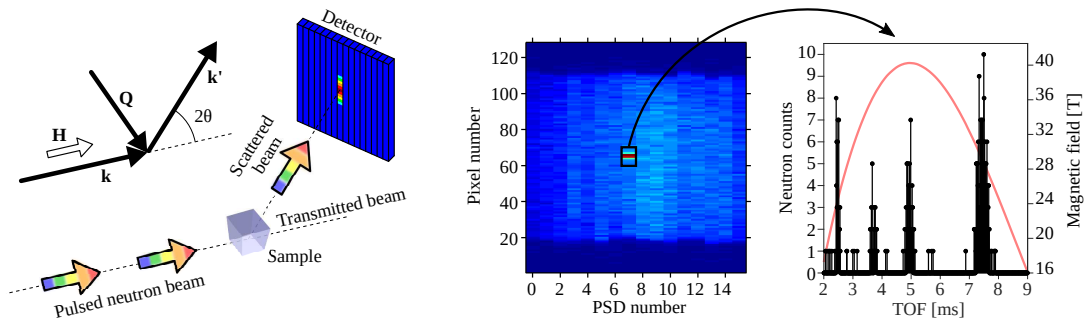}
	\caption{TOF Laue neutron diffraction under pulsed magnetic fields at NOBORU. Scattered neutrons are recorded using a 2D detector consisting of 16 tubes with 128 pixels each. The detector image shows accumulated data for 99 magnet pulses. Integrating neutron counts in the box yields data as shown in the histogram and with the red curve being the magnetic field pulse. At $40\,\mathrm{T}$ there is a signal around $5\,\mathrm{ms}$ which corresponds to the $(0,1,0)$ Bragg peak. The shown data is collected with the settings $\mu_0 H_{\mathrm{max}} = 40\,\mathrm{T}$ and $\Delta t = 1000\,\mathrm{\mu s}$.}
	\label{fig:NOBORUsetup}
\end{figure*}

Whereas measurements of bulk properties such as magnetization and specific heat in magnetic fields approaching $100\,\mathrm{T}$ \cite{takeyama2010,zherlitsyn2012,nguyen2016} are now becoming routine, neutron scattering experiments in magnetic fields greater than $17.3\,\mathrm{T}$ were until recently not possible. The pursuit of higher fields for neutron experiments have followed two different paths. Using pulsed field technology, maximum fields greater than $40\,\mathrm{T}$ can be reached at the price of a very low duty cycle \cite{yoshii2009,nojiri2011}. In comparison, hybrid magnet technology combining superconducting and resistive coils permits continuous operation, but limits the maximum field to $26\,\mathrm{T}$ \cite{smeibidl2016,prokhnenko2015,prokhnenko2016,prokhnenko2017}.

In the present work, pulsed-field magnetization and electric polarization measurements were performed at the Institute for Solid State Physics, Japan. Magnetic field pulses of $40\,\mathrm{ms}$ duration and a peak field of $56\,\mathrm{T}$ were applied along $c$. The magnet coil was made from a copper-silver alloy. A $2\times 2\times 2\,\mathrm{mm^3}$ crystal of spherical shape was used for the magnetization measurements and the absolute value of the magnetization was scaled to previous results obtained with static fields \cite{toftpetersen2011}. A plate-shaped crystal with area $2\times 1\,\mathrm{mm^2}$, thickness $0.65\,\mathrm{mm}$ and $a$ perpendicular to the plate was used for measuring the electric polarization along $a$, $P_a$, using a procedure similar to that described in Refs. \onlinecite{akaki2012,mitamura2007}.

Magnetic structures were directly probed using time-of-flight (TOF) Laue neutron diffraction. In this method, a polychromatic neutron beam is incident on the sample and diffracted beam intensities are recorded at different scattering vectors or momentum transfers, $\bf Q$, by knowing the neutron flight time, $\mathrm{TOF}$, and the travelled distance, $L$, at the detector position. A number of corrections are generally needed in order to convert from the collected integrated neutron intensities, $I ({\bf Q})$, to structure factors \cite{schultz1982}:
\[
	I ({\bf Q}) \propto \Psi (\lambda) \ \epsilon (\lambda) \ |F ({\bf Q})|^2 \ \left( \frac{\lambda^4}{\sin^2 \theta} \right).
\]
Here $\Psi (\lambda)$ is the neutron flux as measured by an upstream monitor, $\epsilon (\lambda)$ is the detector efficiency accounted for by a vanadium measurement, $F ({\bf Q})$ is the structure factor and $\frac{\lambda^4}{\sin^2 \theta}$ is the Lorentz factor with $2\theta$ the scattering angle. In addition, one may consider other factors such as absorption or extinction. For magnetic scattering, additional corrections are needed for the form factor squared, $f(Q)^2$, and for taking into account the relative orientation of $\bf Q$ and the magnetic moments.

One such TOF neutron diffraction experiment was performed at the high magnetic field facility for neutron scattering at the Helmholtz-Zentrum Berlin. The setup consists of the Extreme Environment Diffractometer (EXED) and the High Field Magnet (HFM) \cite{smeibidl2016,prokhnenko2015,prokhnenko2016,prokhnenko2017}. The unique horizontal hybrid solenoid magnet allowed for probing all magnetic phases up to $25.1\,\mathrm{T}$ DC field.
The magnet has a $30^{\circ}$ conical opening, which combined with magnet rotation with respect to the incident neutron beam gives access to a substantial region of reciprocal space. The sample was a high-quality $330\,\mathrm{mg}$ single crystal oriented with $(0,1,0)$ and $(0,0,1)$ in the horizontal scattering plane. Magnetic fields were applied along the $c$-axis with temperatures in the interval $1.3-30\,\mathrm{K}$. The magnet was rotated $-6^{\circ}$ with respect to the incoming beam with wavelength band $0.7-6.9\,\mathrm{Å}$. A number of Bragg peaks were observed on the forward and backscattering area detectors: $(\pm 1, K, 0)$, $(-2, -2, 0)$, $(0, K, 0)$, $(2, -1.33, 0)$, $(1, -0.67, 0)$, $(-2, -0.33, 0)$, $(0, 0, 4)$ and $(0, 0, 2)$ with $K \in \left[ -2, 0 \right]$.

\begin{figure*}
	\centering
	\includegraphics[width = 0.95\textwidth]{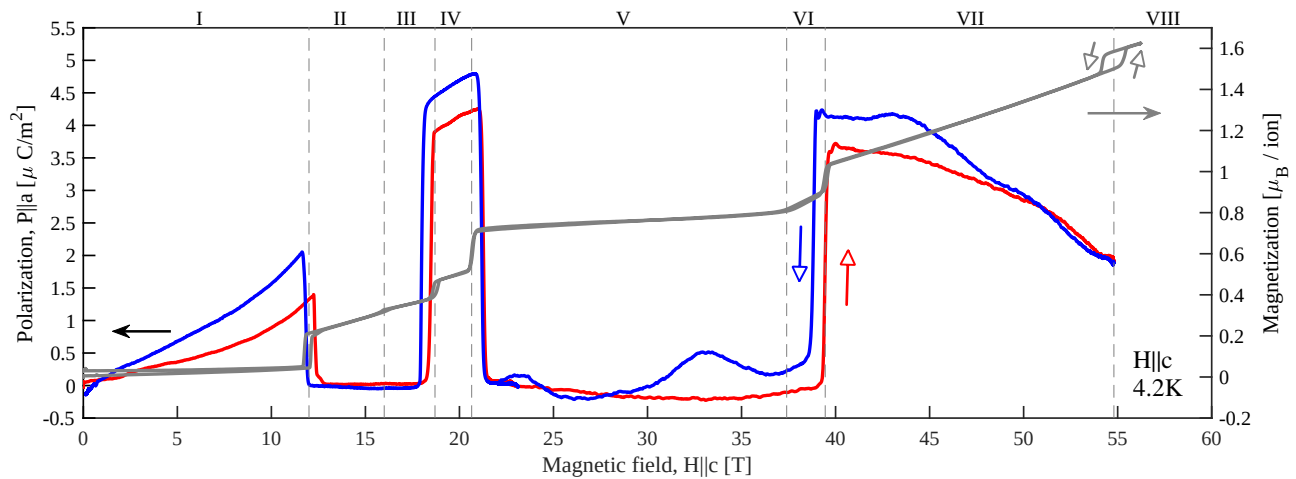}
	\caption{Magnetization (grey curve) and electric polarization data (red and blue curves). Magnetization (right axis) and electric polarization (left axis) along $a$ measured at $4.2\,\mathrm{K}$ as a function of magnetic field applied along $c$. Phase transitions as observed in the magnetization are indicated with vertical dashed lines and the ramp direction is shown with open arrows. Phase numbers are listed on top of the plot.}
	\label{fig:mag_pol}
\end{figure*}

A second TOF neutron diffraction experiment was performed on the NeutrOn Beamline for Observation \& Research Use (NOBORU) at the Japan Proton Accelerator Research Complex. The instrument was operated in Laue mode with wavelengths $\lambda < 10.5\,\mathrm{Å}$ and an area detector (with 16 vertical PSDs of 128 pixels with $10\,\mathrm{\mu s}$ time bins) was placed in forward scattering positions. The pulsed magnetic field was generated by a copper coil mounted on an insert for a standard $^4$He cryostat controlling the sample temperature. The coil itself was immersed in liquid nitrogen and connected to a capacitor bank delivering $10\,\mathrm{ms}$ pulses and thereby generating fields up to $42\,\mathrm{T}$. The sample was the same $330\,\mathrm{mg}$ single crystal also used in the HFM/EXED experiment. It was oriented with the $a$-axis vertical and the $c$-axis in the scattering plane rotated $6^{\circ}$ away from the field direction in order to reach momentum transfers along $(0,K,0)$. This particular direction in reciprocal space was chosen since magnetic structures in the lithium orthophosphates have so far without exception been found to propagate along $b$ \cite{jensen2009_2,toftpetersen2017,fogh2017}. The time delay, $\Delta t$, between neutron pulse and magnet pulse as well as the maximum field, $\mu_0 H_{\mathrm{max}}$, were adjusted such that the neutron TOF-dependent intensity collected in a small region on the area detector may be converted to intensity versus $(0,K,0)$. The relation between TOF and $K$ goes as $K = \frac{2 \alpha \, L \, b \sin \theta}{\mathrm{TOF}}$ with $\alpha = 252.7\,\mathrm{\mu s/m/Å}$. Data was collected using 14 different settings of $\mu_0 H_{\mathrm{max}}$ and $\Delta t$ and with 50-120 magnet pulses per setting. The experimental technique is also described in Refs. \onlinecite{toftpetersen2017} and \onlinecite{yoshii2009}. The setup is illustrated in Fig. \ref{fig:NOBORUsetup}.

\section{Results \& discussion}

\subsection{Magnetometry and electric polarization}

Magnetic phase transitions are observed in the magnetization at $12.0$, $16.0$, $19.1$, $20.9$, $37.6$, $39.4$ and $54\,\mathrm{T}$ as shown in Fig. \ref{fig:mag_pol}. Phases are enumerated using Roman numerals I-VIII, following the notation of Ref. \onlinecite{toftpetersen2017}. Note that the material is magnetized by $\sim \frac{3}{4}$ of the saturation magnetization ($M_S = 2.2\,\mathrm{\mu_B}$/ion \cite{jensen2009_2}) at the highest probed field strength. To our best knowledge, the transitions at $37.6$, $39.4$ and $54\,\mathrm{T}$ have not been reported earlier and phases VI, VII and VIII are unknown. Pronouced hysteresis of about $1\,\mathrm{T}$ is observed in the magnetization at the $54\,\mathrm{T}$ transition but not at $37.6$ and $39.4\,\mathrm{T}$.

Figure \ref{fig:mag_pol} also shows the electric polarization along $a$ as a function of magnetic field applied along $c$. Phases I, IV and VII display the ME effect with finite tensor elements, $\alpha_{ac}$ and/or $\beta_{acc}$. Hysteresis in the polarization is observed at $12.0$, $19.1$ and $39.4\,\mathrm{T}$ but not at $20.9\,\mathrm{T}$. The bumps observed in the polarization around $\sim 23,\sim33$ and $\sim44\,\mathrm{T}$ for decreasing fields are attributed to mechanical oscillations of the sample and probe in the experimental setup. Because of the delayed response, these disturbances often appear only for decreasing field as seen here.

Previously, the ME effects in phases I and IV have been studied in Refs. \onlinecite{mercier,toftpetersen2017,khrustalyov2016_Ni}. A model for the temperature dependence of $P_a$ in phase I was put forth Ref. \onlinecite{jensen2009_2}. Expanding on this theoretical framework, a similar model for the field-dependence of $P_a$ in phased IV was developed in Ref. \onlinecite{toftpetersen2017}. In Section III.D we will explore if this model can explain the field-dependence of $P_a$ in phase VIII as well, despite the pronounced differences in characteristics that can be summarized as follows: in phase I, $P_a$ is approximately linear with field until around $6.5\,\mathrm{T}$ where a quadratic onset is evident \cite{toftpetersen2017} (we note that Ref. \onlinecite{khrustalyov2016_Ni} reported the upturn to be cubic in field and not quadratic). In phase IV, $P_a$ is linear for the entire field interval. In both phases I and IV, $\frac{dP_a}{dH}>0$, i.e. $\alpha_{ac}$ is positive and the quadratic tensor element, $\beta_{acc} > 0$, is also positive above $6.5\,\mathrm{T}$ in phase I. In phase VII, however, $P_a$ appears purely quadratic and $\frac{dP_a}{dH}<0$, i.e. $\alpha_{ac} \approx 0$ and $\beta_{acc} < 0$.

Before further discussing the ME effect in phase VII, we first describe our neutron diffraction experiments in Sections III.B and C.



\subsection{Neutron diffraction}



\begin{figure}
	\centering
	\includegraphics[width = 0.95\columnwidth]{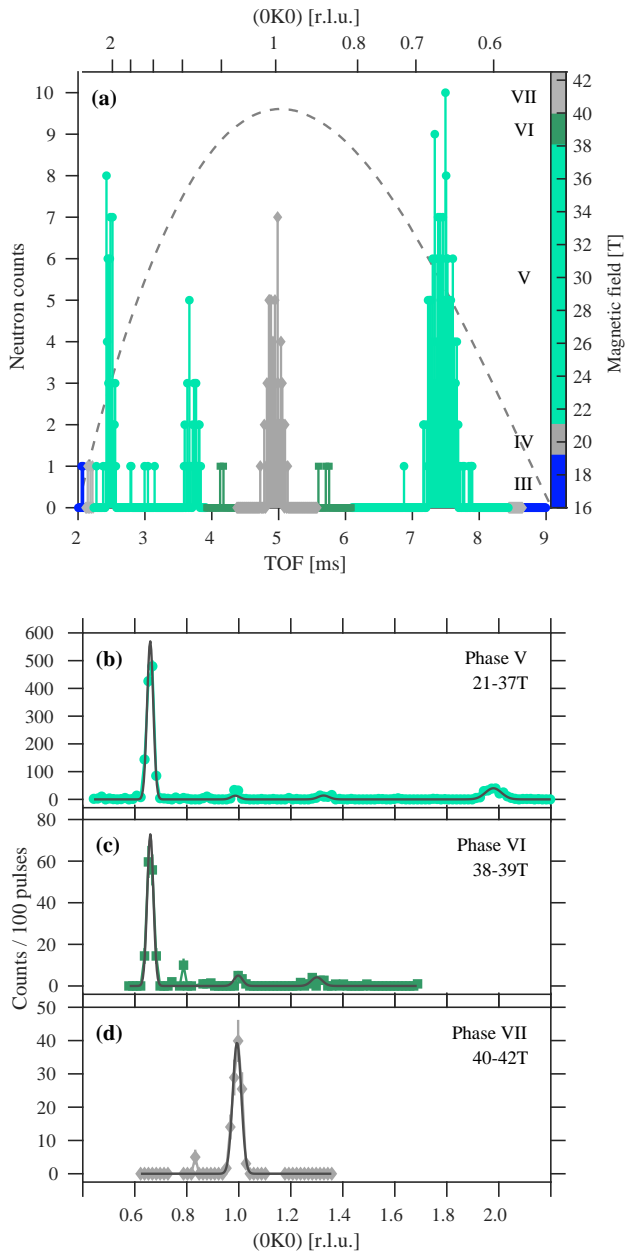}
	\caption{Pulsed-field Laue neutron diffraction. (a) Example of accumulated raw data for 99 pulses at one magnet setting: $\mu_0 H_{\mathrm{max}} = 40.5\,\mathrm{T}$, $\Delta t = 1000\,\mathrm{\mu s}$. Neutron counts are shown as a function of TOF [bottom axis] as well as corresponding scattering vector, $(0,K,0)$ [top axis]. The colors represent the field intervals in which each neutron has been detected. The magnetic field pulse is shown by the dashed line in the background with field values read to the right of the colorbar. Note that the scale starts at $16\,\mathrm{T}$. For clarity, errors of the neutron data are not shown but are simply $\sqrt{N}$ Poisson counting errors. Panels (b)-(d) show integrated neutron counts for each of the phases V, VI and VII as a function of $(0,K,0)$ for all the data collected. The error bars show the propagated error and the bin size is $\Delta K = 0.015\,\mathrm{r.l.u.}$ The solid lines show Gaussian fits to the observed peaks. Note that there is only data shown at positions of $(0,K,0)$ that have been probed in the experiment, e.g. $(0,2,0)$ was not probed at fields above $37\,\mathrm{T}$.}
	\label{fig:NOBORU}
\end{figure}

\begin{figure*}
	\includegraphics[width = \textwidth]{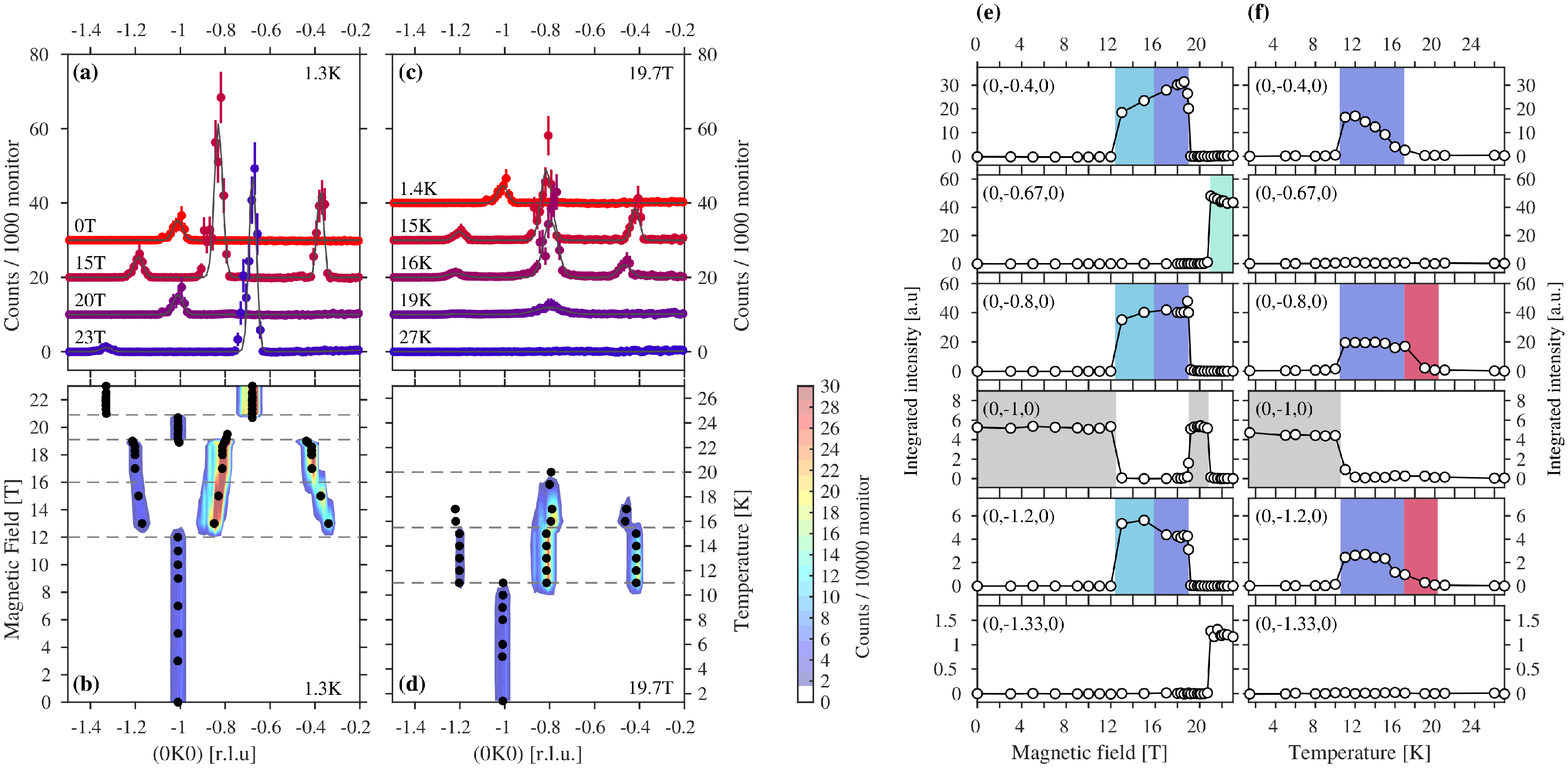}
	\caption{Temperature and field dependencies of the $(0,K,0)$ Bragg peak. Top lefthand panels, (a) and (c), show neutron intensity profiles as a function of $(0,K,0)$ at selected field values at $1.3\,\mathrm{K}$ and at selected temperatures at $19.7\,\mathrm{T}$ respectively. Data sets are offset on the vertical axis for clarity. Gaussian profiles were fitted to the line shapes (solid lines) and the integrated intensity calculated. Panels (b) and (d) show the neutron intensities in a color plot with scattering vector position, $(0,K,0)$, and field or temperature on the axes. Fitted positions are marked with black dots. Horizontal dashed lines indicate phase transitions. Righthand panels, (e) and (f), show integrated intensities for the identified scattering vectors: $(0,-1,0)$, $(0,-1 \pm k,0)$ and $(0,-2k,0)$ for $k \approx 0.2$ and $(0,-1 \pm k,0)$ for $k = 0.33$. Different phases are indicated with colored regions corresponding to the phase diagram in Fig. \ref{fig:phasediagram}.}
	\label{fig:HFM-EXED}
\end{figure*}

A pulsed-field neutron diffraction experiment was performed at NOBORU as described in Section II and an example of the raw data is shown in Fig. \ref{fig:NOBORU}(a). Four distinct peaks are observed at $2.5$, $3.7$, $5.0$ and $7.5\,\mathrm{ms}$ corresponding to momentum transfers $(0,2,0)$, $(0,\frac{4}{3},0)$, $(0,1,0)$ and $(0,\frac{2}{3},0)$ respectively. The nuclear peak, $(0,2,0)$, is present at all fields whereas the remaining peaks are magnetic and only appear in specific field intervals. The Bragg peak $(0,1,0)$ is observed in phase VII whereas $(0,\frac{4}{3},0)$ and $(0,\frac{2}{3},0)$ are present in phases V and VI. Below $2\,\mathrm{ms}$ (not shown), the spectrum is dominated by background counts originating from high-energy particles but at higher TOFs the background is extremely low: 0-1 counts per 100 pulses. 

Figures \ref{fig:NOBORU}(b)-(d) show the integrated intensities for the field intervals $21-37\,\mathrm{T}$ (phase V), $38-39\,\mathrm{T}$ (phase VI) and $>40\,\mathrm{T}$ (phase VII), respectively. The intervals are chosen with approximately $\pm 0.5\,\mathrm{T}$ distance to the phase boundaries obtained from the magnetization measurements. Due to the rapidly varying field this was done as a precaution in order to exclusively sum up neutrons scattered while the field was well away from the phase boundaries. In phase V, a strong peak is observed at $(0,\frac{2}{3},0)$ as well as weaker ones at $(0,1,0)$, $(0,\frac{4}{3},0)$ and $(0,2,0)$. The situation is similar in phase VI with a strong peak at $(0,\frac{2}{3},0)$ and weaker ones at $(0,1,0)$ and $(0,\frac{4}{3},0)$. Finally, in phase VII, the peaks at $(0,\frac{2}{3},0)$ and $(0,\frac{4}{3},0)$ give way to a sole peak at $(0,1,0)$. Note that $(0,2,0)$ was not probed in phases VI and VII.

Peak positions were obtained from fits to Gaussian profiles. The peak widths were fixed based on analysis of zero-field data which displayed nuclear peaks $(0,K,0)$ with $K = 2, 4, 6, 8, 10$ and magnetic peaks with $K = 1, 3$. These data (not shown) are of much higher statistical quality than the pulsed field data, and allow us to reduce the number of fitting parameters and thereby obtain stable fits. For $K < 6$ the peak widths approximately follow a linear trend: $\sigma(K) = \alpha K + \beta$, where $\alpha = 0.0143(1)$ and $\beta = 0.0023(7)\,$r.l.u. were fitted. This relation is used for fixing the peak widths in the field-on data. The fitted peak positions in phase V are $(0,0.6593(6),0)$, $(0,0.987(4),0)$, $(0,1.326(5),0)$ and $(0,1.979(3),0)$. In phase VI they are similarly $(0,0.660(1),0)$, $(0,1.00(1),0)$ and $(0,1.300(8),0)$. In phase VII a single peak is observed at $(0,0.993(2),0)$. Note that the propagation vectors are assumed field independent for each individual phase. While this is experimentally verified up $23\,\mathrm{T}$ for phase V [see colorplot in Fig. \ref{fig:HFM-EXED}(b)], it is an assumption at all higher fields.


The pulsed-field technique is limited by counting statistics since the setup has a $10-30\,\mathrm{min}$ cool-down period after each magnet pulse in which no data is collected. It is therefore impractical for detailed studies of phase boundaries. The HFM/EXED facility, on the other hand, is excellent for parametric studies and allowed for tracking magnetic phase boundaries in LiNiPO$_4$ up to $25.1\,\mathrm{T}$. Examples of collected data are shown in Fig. \ref{fig:HFM-EXED}. Moreover, the superior counting statistics at EXED enabled improved peak position determination yielding $(0,-1.009(1),0)$ for phase IV as well as $(0,-1.331(9),0)$, and $(0,-0.68(1),0)$ for phase V respectively. These positions correspond to propagation vectors ${\bf Q}_{\mathrm{IV}} = (0, 0, 0)$ and ${\bf Q}_{\mathrm{V}} = (0, \frac{1}{3}, 0)$ as also previously proposed \cite{toftpetersen2017}. It is pointed out that the exact values of $k = 0$ and $\frac{1}{3}$ are conjectured within the experimental resolution. It is however possible that $k \neq 0$ in phase IV but that the period of the magnetic structure is very long and therefore almost matches the nuclear cell. Likewise, the period of phase V may not be exactly 3 crystallographic unit cells. On the other hand, the scattering vector in phase I -- which is known to be commensurate \cite{jensen2009_2} -- is determined to ${\bf Q}_{\mathrm{I}} = (0,-1.009(1),0)$. This is precisely the same as in phase IV. We therefore maintain that phase IV is truly commensurate.

In order to obtain the intensity at a certain $(0,K,0)$ position, neutron counts were summed in slices of thickness $H \in \left[-0.1,0.1 \right]$ and $L \in \left[-0.05,0.05 \right]$. Subsequently, Gaussian profiles were fitted to the line shapes of neutron counts as a function of $K$ and the integrated intensities calculated. No vanadium or Lorentz corrections were applied here since only phase transitions were of interest and not absolute intensities. However, note that the Lorentz factor accounts for the higher intensities in Fig. \ref{fig:HFM-EXED} at shorter $Q$ corresponding to higher values of $\lambda$ (or equivalently, longer TOFs). The paramagnetic background at $0\,\mathrm{T}$, $43\,\mathrm{K}$ was subtracted for all data sets.

Figures \ref{fig:HFM-EXED}(a), (b) and (e) present results from a field scan performed at $1.3\,\mathrm{K}$. Intensity appears at peak positions $(0,-1, 0)$, $(0,-1 \pm k,0)$ and $(0,-2k,0)$ with the value of $k$ depending on the field. The $(0,-1,0)$ reflection is present for $0-12.5(5)\,\mathrm{T}$ as well as for $19.0(1)-20.9(2)\,\mathrm{T}$. Peaks with $k \approx 0.2$ are characteristic of the spiral phase and appear in the interval $12.5(5)-19.0(1)\,\mathrm{T}$ with the 5 unit cell period lock-in at $16\,\mathrm{T}$. Above $20.9(2)\,\mathrm{T}$, neutron intensity is observed at $k = \frac{1}{3}$. These observations are in excellent agreement with previous results \cite{jensen2009_2,kharchenko2010,toftpetersen2011,toftpetersen2017}. No hysteresis was observed at the transitions at $19.0(1)$ and $20.9(2)\,\mathrm{T}$.

The results of a temperature scan at $19.7\,\mathrm{T}$ are shown in Figs. \ref{fig:HFM-EXED}(c), (d) and (f). The linearly modulated phase, spiral phases and the commensurate phase IV are encountered in succession upon cooling. Intensity is observed at positions $(0,-1 \pm 0.2,0)$ in the linearly modulated phase starting around $20-21\,\mathrm{K}$. At $16\,\mathrm{K}$, the position of the vector changes towards longer magnetic unit cell periods 
characterizing the spiral phases. The incommensurate peaks give way to $(0,-1,0)$ around $10\,\mathrm{K}$ when finally entering phase IV.

\begin{figure}[t!]
	\centering
	\includegraphics[width = 0.95\columnwidth]{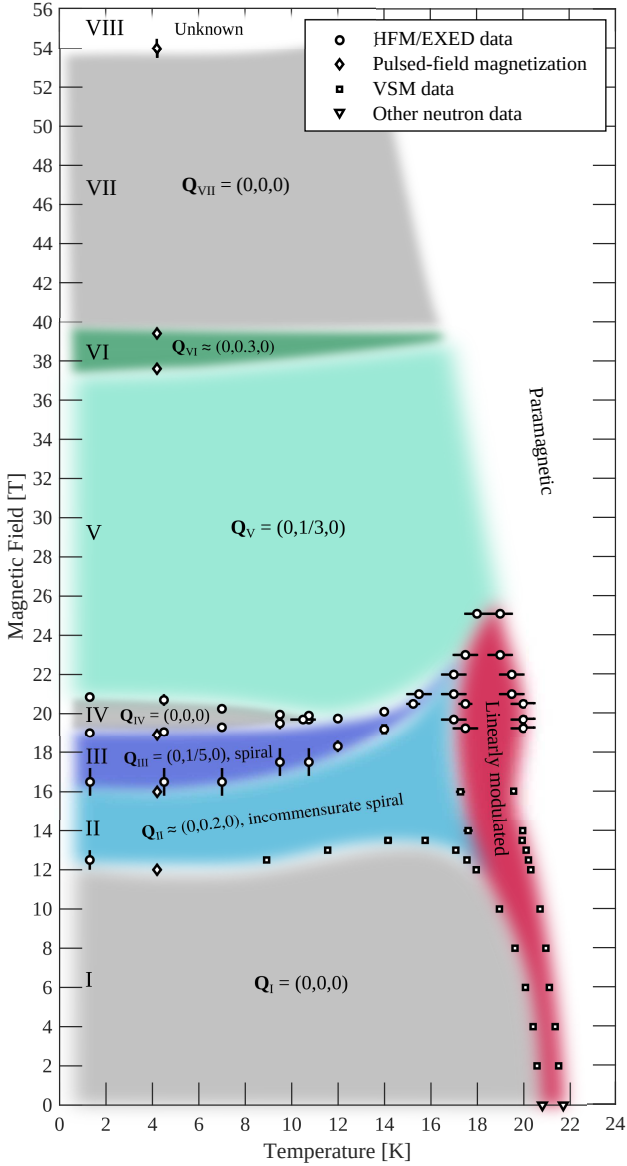}
	\caption{Magnetic phase diagram of LiNiPO$_4$ based on neutron diffraction, magnetization measurements in pulsed and static fields as well as pulsed field electric polarization measurements. In all cases the field was along the crystallographic $c$-axis. The error bars of the vibrating sample magnetometry (VSM) data are comparable to the symbol size. The three phases with propagation vector $(0,0,0)$ support the ME effect (grey regions) whereas all phases with larger periods do not (colored regions). The field-induced phases are enumerated I-VIII for increasing field. Note that the phase boundaries for temperatures $T>4.2\,\mathrm{K}$ and fields $\mu_0H > 25.1\,\mathrm{T}$ have not been probed and the boundaries indicated here are merely a conjecture.}
	\label{fig:phasediagram}
\end{figure}

Thus following the $(0,K,0)$ magnetic Bragg peak as a function of temperature and magnetic field as measured at HFM/EXED together with magnetization and polarization measurements enable the determination of the magnetic phase diagram of LiNiPO$_4$ up to $56\,\mathrm{T}$. The result is shown in Fig. \ref{fig:phasediagram}. The magnetic structures in phases IV-VII are discussed in the next section.

\subsection{Magnetic structures}

\begin{figure}[t!]
	\centering
	\includegraphics[width = \columnwidth]{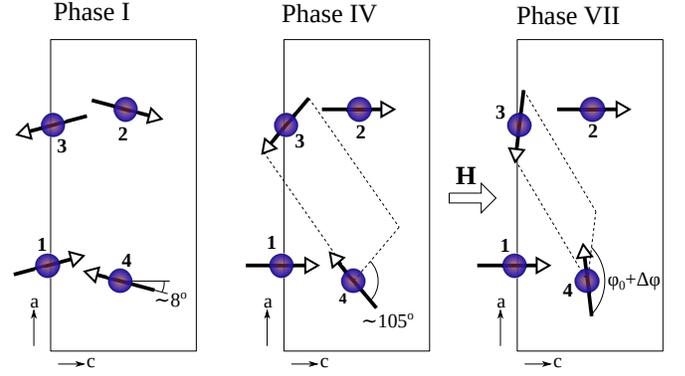}
	\caption{Magnetic structures in the ME phases I, IV and VII projected to the $(a,c)$-plane. The zero-field canting angle \cite{jensen2009_2} is shown in phase I with this highly symmetric configuration valid at low fields only. In phase IV and VII, spins 1 and 2 are aligned with the applied magnetic field and spins 3 and 4 are canted. The angle shown here between spins 3 and 4 in phase IV is for just upon entering the phase.}
	\label{fig:magneticstructures}
\end{figure}

A magnetic structure in phase IV was proposed in Ref. \onlinecite{toftpetersen2017} based on the observation of a single magnetic Bragg peak -- $(0,1,0)$ -- together with magnetization data. A model for the ME effect further substantiated the proposed commensurate structure consisting of a $(\uparrow \uparrow \downarrow \downarrow)$ symmetry component along $c$ as well as two equally large components of $(\uparrow \downarrow \downarrow \uparrow)$ and $(\uparrow \downarrow \uparrow \downarrow)$ symmetry respectively, both with spins polarized along $a$. The structure is illustrated in Fig. \ref{fig:magneticstructures}.

Apart from parametric studies of the $(0,K,0)$ magnetic Bragg peak, the HFM/EXED experiment also allowed for the observation of additional magnetic Bragg peaks in phase IV. In addition to the $(0,1,0)$ peak already observed in our previous pulsed field experiment, magnetic intensity was thus observed at $(\pm 1, -2, 0)$ and $(\pm 1, -1, 0)$ in phase IV at HFM/EXED. Those peaks represent structure components $(\uparrow \downarrow \downarrow \uparrow)$ and $(\uparrow \downarrow \uparrow \downarrow)$, respectively, and with spin mostly oriented along $a$. Hence, the additional magnetic peaks are consistent with the structure proposed in Ref. \onlinecite{toftpetersen2017}.

Neutron diffraction, magnetization data and mean-field theory presented in Ref. \onlinecite{toftpetersen2017} lead to a proposal for the structure in phase V where the propagation vector is $(0,\frac{1}{3},0)$, i.e. a period of 3 crystallographic unit cells along $b$. This spin structure consists of a ferromagnetic component along $c$ and an antiferromagnetic symmetry component $(+ + -\beta -\beta)$ describing the modulated part of the structure. Here the notation is slightly altered like in Ref. \onlinecite{jensen2009_2} such that $+$ ($-$) denotes spin $\uparrow$ ($\downarrow$) and $\beta = e^{-i \pi /3}$ is a phase factor.

The structure proposed in Ref. \onlinecite{toftpetersen2017} was based on the observation of the $(0,\frac{4}{3},0)$ magnetic Bragg peak and a magnetization which exhibits a near-plateau at $\frac{1}{3}$ saturation magnetization. A number of additional magnetic Bragg peaks were observed at the HFM/EXED experiment and a somewhat sounder structure determination is in principle possible. Intensities are obtained using the Mantid software package \cite{arnold2014} as follows: (1) rectangular masks are created for each individual peak, (2) a second order polynomial is fitted to empirically describe the background of the TOF spectrum. Next, (3) the background is subtracted and finally, (4) vanadium and Lorentz corrections are applied. Due to technical issues not all the intensities could be reliably determined. E.g. the equivalent peaks $(\pm 1, -\frac{4}{3},0)$ differ by a factor of $\sim1.5$ and $(-2,-\frac{1}{3},0)$ is placed near the edge of the detector. Still, the presence or absence of these additional peaks may in the least be used in the analysis. In an attempt to determine the magnetic structure in phase V, intensities for a number of model structures were refined using \textsc{Fullprof} \cite{rodriguezcarvajal1993} and compared with the observed intensities [see Table \ref{tab:LiNiPO4_HFM-EXED_incommensurate}]. The models count a spin-density wave with spins along $c$ as well as circular and elliptical spiral structures with spins in the $(a,c)$-plane. The elliptical spiral has major axis along $c$. The propagation vector is $(0,\frac{1}{3},0)$ in all cases. The spin-density wave forbids neutron intensity for $(1,-\frac{2}{3},0)$ and $(\pm 1, -\frac{4}{3},0)$. Since these peaks are present, this model can readily be discarded. The circular and elliptical spiral structures both allow all observed Bragg peaks. Although the data quality does not allow for a conclusive distinction between the two, the circular spiral yields a better refinement.

\begin{table}[t!]
	\caption{Observed vs. calculated magnetic intensities for structures proposed in phase V. }
	\label{tab:LiNiPO4_HFM-EXED_incommensurate}
	\begin{ruledtabular}
	\begin{tabular}{c | c | c c c c}
		$(H,K,L)$		& Obs. int.	& Linear		& Circular	& Elliptical\\
		\hline
		$(0,-2/3,0)$		& 11.218(6)	& 11.38		& 11.36		& 18.38\\		
		$(1,-2/3,0)$		& 0.526(9)	& 0			& 0.66		& 0.99\\		
		$(-2,-1/3,0)$	& 3.76(33)	& 0.96		& 1.94		& 3.10\\		
		$(0,-4/3,0)$		& 13.26(4)	& 5.44		& 5.43		& 8.78\\
		$(-1,-4/3,0)$	& 0.155(7)	& 0			& 0.39		& 0.59\\
		$(1,-4/3,0)$		& 0.112(16)	& 0			& 0.39		& 0.59\\
		$(2,-4/3,0)$		& 24.5(3.1)	& 4.28		& 4.50		& 7.29\\
	\end{tabular}
	\end{ruledtabular}
\end{table}

Phase VI looks very similar to phase V [compare Figs. \ref{fig:NOBORU}(b) and (c)]. Yet, the magnetic susceptibility, $\frac{dM}{dH}$, is a factor $\sim 10$ larger in phase VI compared to phase V [re-visit the magnetization curve in Fig. \ref{fig:mag_pol}]. Furthermore, the period of the structure is possibly longer with a peak observed at $(0, 1.300(8), 0)$ in phase VI as compared to $(0, 1.326(5), 0)$ in phase V. When determining the peak position it was assumed field-independent but as also previously pointed out, this might not be the case. If e.g. $K$ decreases with field from $K = 1.33$ to $K = 1.27$ within the field interval, the fitted position -- given that the neutron intensity stays constant -- would indeed be $K = 1.30$. In such a case, the period of the magnetic structure would no longer be locked in with the crystal structure. However, if the peak is actually moving with field, a peak broadening is expected when integrating over the entire field interval. This does not appear to be the case when inspecting Fig. \ref{fig:NOBORU}(c).


Having thus described phases IV, V and VI, we now turn to phase VII. In many ways, this phase looks similar to phase IV: the magnetization is linear as a function of applied field [see Fig. \ref{fig:mag_pol}] and a single magnetic Bragg peak -- $(0, 1, 0)$ -- was observed in the pulsed-field Laue neutron diffraction experiment. The magnetization is $\sim 1.1\,\mathrm{\mu_B} = \frac{1}{2} M_S$ $(M_S = 2.2\,\mathrm{\mu_B}$ for LiNiPO$_4$ \cite{jensen2009_2}) at the phase transition at $H_c = 39.4\,\mathrm{T}$. This may be obtained by a further magnetized version of the structure in phase IV. In the proposed structure, spins 1 and 2 are aligned with the applied magnetic field and spins 3 and 4 are almost antiparallel to each other as well as perpendicular to the field [see Fig. \ref{fig:magneticstructures}]. The angle between spins 3 and 4 is $\varphi_0 \approx \pi$ upon entering phase VII and increases, $\varphi_0 + \Delta \varphi > \pi$, as the field is increased. 

Finally, the presented data is insufficient to comment on the likely magnetic structure in phase VIII ($>54\,\mathrm{T}$). Further work along this direction will have to await further developments in pulsed-field technology for neutron diffraction.

To summarize this section on magnetic structures, the magnetic phase diagram of LiNiPO$_4$ is presented in Fig. \ref{fig:phasediagram}. It consists of a series of alternating commensurate and incommensurate phases. Strikingly, all the observed ${\bf Q} = (0,0,0)$ phases display the ME effect and all phases with larger periods do not. In the next section we will have a closer look at the quadratic ME effect discovered in phase VII.

\subsection{Quadratic magnetoelectric effect}

\begin{figure}
	\centering
	\includegraphics[width = \columnwidth]{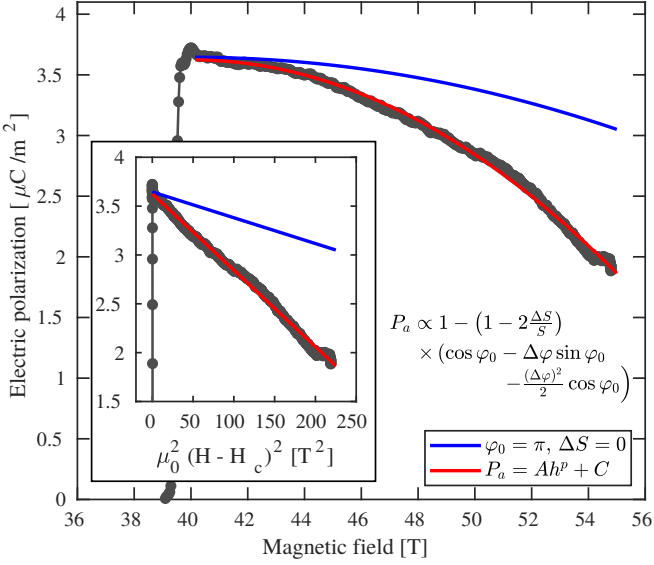}
	\caption{Electric polarization as a function of field in phase VII with the blue line showing the polarization as a function of field as calculated using the model described in the text with $\varphi_0 = \pi$ and $\Delta S = 0$. The red line shows a fit to a general function, $P_a = A h^p + C$, also described in the text. The inset shows that the polarization is close to linear as function of the reduced field squared.}
	\label{fig:MEeffect}
\end{figure}

As already mentioned, a magnetic-field-induced electric polarization is observed in phases I, IV and VII [re-visit Fig. \ref{fig:mag_pol}], precisely those phases with propagation vector $(0,0,0)$ and where the magnetic unit cell is identical to the crystallographic one. In all three cases, the measured polarization, $P_a$, is triggered by a magnetic field applied along $c$. Thus, the non-zero ME tensor elements are $\alpha_{ac}$ or $\beta_{acc}$. However, as also pointed out in Section III.A, the field-dependencies of these tensor elements are different in phase VII as compared to phases I and IV. In phase I the linear ME tensor element is $\alpha_{ac} > 0$ and there is an onset of a second order effect around $6.5\,\mathrm{T}$ with $\beta_{acc} > 0$. In phase IV, $\alpha_{ac} > 0$ and $\beta_{acc} \approx 0$. In phase VII, however, the linear effect is entirely replaced by the quadratic effect and $\alpha_{ac} \approx 0$, $\beta_{acc} <0$. This is demonstrated in the inset in Fig. \ref{fig:MEeffect} where the electric polarization is plotted as a function of the reduced field, $h = \mu_0 (H - H_c)$, squared. It is also noteworthy that the quadratic ME tensor element has opposite sign in phase VII as compared to phase I. Since $\beta_{acc} < 0$, $\alpha_{ac} \approx 0$ but $P_a > 0$ in phase VII, a constant term, $P_0 > 0$, must exist. This means that phase VII is not only ME but in some sense also pyroelectric.

The appearance of both the linear and second order ME effect is governed by the magnetic symmetry of the crystal. The magnetic point group of LiNiPO$_4$ in phase I is $mm'm$ which allows linear ME coefficients $\alpha_{ac},\alpha_{ca} \neq 0$ but the quadratic effect is prohibited \cite{grimmerInternationalTablesD}. The proposed magnetic structures in phases IV and VII lead to the magnetic point group $2'm'm$ and now both linear and quadratic ME effects are allowed with tensor elements $\alpha_{ac}, \alpha_{ca} \neq 0$ and $\beta_{aaa}, \beta_{abb}, \beta_{acc}, \beta_{bba}=\beta_{bab}, \beta_{cca}=\beta_{cac} \neq 0$ \cite{grimmer1994}. Thus, the observed non-zero elements $\alpha_{ac}$ in phase IV and $\beta_{acc}$ in phase VII are consistent with the magnetic point group of the proposed spin structures. Moreover, the magnetic point group in phase I also becomes $2'm'm$ upon applying a magnetic field as an asymmetry in the canting angles is introduced, i.e. in Fig. \ref{fig:magneticstructures} spins 1 and 2 experience a decrease in canting angle whereas spins 3 and 4 obtain a larger canting angle. At low fields, the deviation from $mm'm$ is negligible but starting at $\sim6.5\,\mathrm{T}$ a non-linear response in the electric polarization is clearly seen. It should be mentioned that the branching away from the linear behavior in phase I was already reported in Ref. \onlinecite{khrustalyov2016_Ni}. There, it was assumed that the point group remains $mm'm$ and hence the quadratic term is prohibited. Instead, a cubic term is possible which was then used to describe the data in Ref. \onlinecite{khrustalyov2016_Ni}. It is difficult to unambiguously determine whether the curve follows a quadratic or cubic behavior as a function of applied magnetic field. However, the magnetic point group symmetry in phase IV and VII does change and as argued above this change may take place already in phase I. In the following analysis we carry on assuming that $\beta_{acc} \neq 0$ and disregard any possible cubic contributions to the ME response.

The change of sign in $\beta_{acc}$ may be understood by considering a variation of the model that has previously been successful in describing the field dependence of the induced electric polarization in both phases I and IV\cite{toftpetersen2017}. The exchange energy for spin pairs $(1,2)$ and $(3,4)$ is simply $E_0 = J_{12}S^2 + J_{34} S^2 \cos \left( \varphi_0 + \Delta \varphi \right)$. This is upon entering phase VII where $\varphi_0 \approx \pi$, $\Delta \varphi = 0$ and $J_{12} = J_{34} = J$. When increasing the field the spins $3$ and $4$ rotate further and now $\Delta \varphi > 0$. This is assumed to introduce an asymmetry in the exchange interactions such that $J_{12} \rightarrow J - \lambda x$ and $J_{34} \rightarrow J + \lambda x$, where $\lambda$ is a proportionality constant and $x$ is the displacement of the PO$_4$ tetrahedra. Since the spin pairs (1,2) and (3,4) are no longer equivalent, we also introduce the possibility for different thermal averages of the moment, $S_{12} \rightarrow S$ and $S_{34} \rightarrow S - \Delta S$. Now the exchange energy reads $E_1 = (J-\lambda x) S^2 + (J+\lambda x)(S - \Delta S)^2 \cos \left( \varphi_0 + \Delta \varphi \right)$. Ignoring higher order terms in $\Delta S$, the change in exchange energy is then $\frac{\Delta E}{S^2} = \lambda x \left[ -1  + \left( 1 - 2 \frac{\Delta S}{S} - \frac{2 J}{\lambda x} \frac{\Delta S}{S} \right) \cos \left( \varphi_0 + \Delta \varphi \right) \right]$. Moving the PO$_4$ tetrahedra, which in our model is responsible for the occurence of finite polarization, also introduces an elastic energy, $\epsilon_x x^2$. The equilibrium displacement is found by minimizing the overall change in exchange and elastic energies. Expanding the cosine $\cos \left( \varphi_0 + \Delta \varphi \right) \approx \cos \varphi_0 - \Delta \varphi \sin \varphi_0 - \frac{(\Delta \varphi)^2}{2} \cos \varphi_0$ then yields an expression for the electric polarization, $P_a = \mathcal{K} x$, as follows:
\begin{align*}
	P_a &= \mathcal{K} \frac{\lambda}{2 \epsilon_x} \left[ 1 - \left( 1 - 2 \frac{\Delta S}{S} \right) \bigg( \cos \varphi_0 - \Delta \varphi \sin \varphi_0  \right.  \\
	& \hspace{5cm} \left.  - \frac{(\Delta \varphi)^2}{2} \cos \varphi_0 \bigg) \right],
\end{align*}
where $\mathcal{K}$ is a proportionality constant. This simplifies to $ P_a = \mathcal{K} \frac{\lambda}{2 \epsilon_x} \left( 2 - \frac{1}{2} (\Delta \varphi)^2 \right)$ for $\Delta S = 0$ and $\varphi_0 = \pi$ which is close to the value $\varphi_0 = 165^{\circ}$ as deduced from the measured magnetization. Hence, the polarization decreases with $(\Delta \varphi)^2$. It is expected that the change in angle is proportional to the reduced field, i.e. $\Delta \varphi \propto h$, such that the electric polarization decreases quadratically with the reduced field. The quadratic ME coefficient may then be identified as $\beta_{acc} \propto - \frac{\mathcal{K} \lambda}{4 \epsilon_x}$ and $P_0 = \frac{\mathcal{K} \lambda}{\epsilon_x}$. Thus, the observed quantities $\beta_{acc} < 0$, $P_0 > 0$ as well as $\alpha_{ab} = 0$ in phase VII appear naturally as a result of Taylor expanding the cosine function around $\varphi_0 = \pi$.

\begin{figure*}
	\centering
	\includegraphics[width = \textwidth]{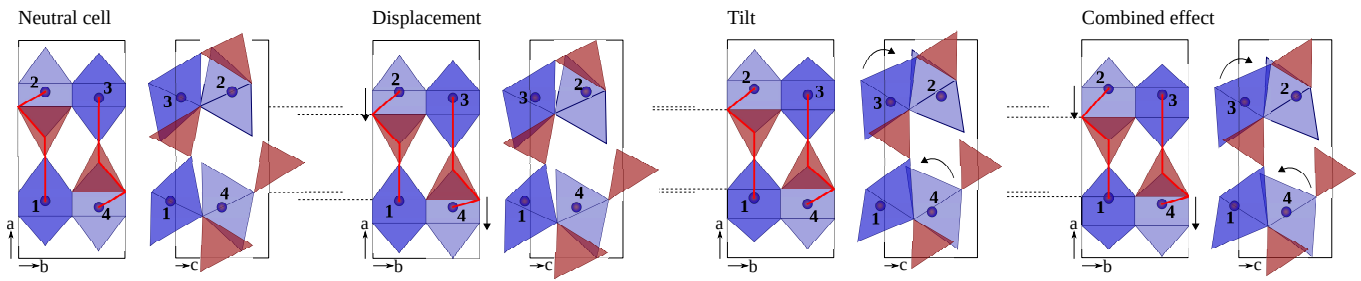}
	\caption{Unit cell sketches showing the effect on the $J_{12}$ and $J_{34}$ exchange paths (red paths) by displacing the PO$_4$ tetrahedra (red regions) along $a$ and/or by tilting the NiO$_6$ (blue regions) octahedra around $b$. Arrows indicate these movements. The positions of the magnetic Ni$^{2+}$ ions (blue spheres) are fixed. The horizontal dashed lines show the oxygen displacement with respect to the neutral cell.}
	\label{fig:MEsketches}
\end{figure*}

The above expression captures the qualitative behavior of $P_a(h)$ [see Fig. \ref{fig:MEeffect}] but the model curve (blue line) does not describe the measured curve well when going to higher magnetic fields and away from the phase transition. Various other attempts to fit the full expression for the electric polarization yields either bad fits, unphysical parameters or results in disagreement with the measured magnetization. Instead, a general phenomelogical function, $P_a = A h^p + C$, with constant parameters $A$ and $C$, yields a fitted exponent $p = 2.01$, i.e. very close to the quadratic behavior obtained by the model. This function describes the data well (red line in Fig. \ref{fig:MEeffect}). 

In summary, although our model captures the qualitative behavior of the observed electric polarization, it lacks some elements in order to give a quantitatively accurate description. Nevertheless, it is remarkable that a model rooted in the same general assumptions can embrace the field dependencies of the induced electric polarization in all three observed ME phases in LiNiPO$_4$. It shows that the ME effect in this compound is governed by a single mechanism which prevails at very high magnetic fields. The simple model is based on varying exchange couplings in certain ways and is as such an empirical description offering little in-depth understanding of the physical phenomenon at work. However, the model does describe our observations well and it is especially encouraging that models emanating from the same base point are able to describe several different ME phases. Therefore, we now speculate on plausible underlying microscopic mechanisms responsible for the ME effect in LiNiPO$_4$.

The magnetic ions are located in a distorted octahedral environment [see Fig. \ref{fig:structure}]. The super exchange bonds involved in the model calculation, $J_{12}$ and $J_{34}$, follow the path Ni-O-P-O-Ni. Two ways of altering the exchange path readily spring to mind: (i) displacing the PO$_4$ tetrahedra along $a$ as also suggested in Ref. \onlinecite{jensen2009_2} or (ii) rotating the NiO$_6$ octahedra around the $b$-axis. The effect on the $J_{12}$ and $J_{34}$ exchange paths of these two mechanism separately and combined are illustrated in Fig. \ref{fig:MEsketches}. Displacing the PO$_4$ tetrahedra changes the bond angles which in turn changes the exchange integrals according to the Anderson-Goodenough-Kanamori rules \cite{goodenough} such that $J_{12} \neq J_{34}$. Furthermore, there is an overall displacement of charge inside the unit cell. This way the exchange integrals and charge displacement are directly coupled and a ME link is created. Similarly, tilting the NiO$_6$ octahedra results in changes in the bond angles and again $J_{12} \neq J_{34}$. However, the oxygen ions are displaced symmetrically such that there is zero net charge displacement and therefore no ME effect. The two mechanisms combined -- PO$_4$ displacement and NiO$_6$ tilting -- yields asymmetric exchange paths as well as asymmetric charge displacements. This enables asymmetric changes in the exchange interactions and in the ordered moment, as proposed in our model.


On a final note, we return to the consequences of the change of magnetic point group symmetry upon applying a magnetic field. In phase I above $6.5\,\mathrm{T}$ and in phases IV and VII, the four Ni sites 1-4 are no longer equivalent but split up in two different sites with spins 1 and 2 on one site and spins 3 and 4 on the other. This means that the crystallographic symmetry is also reduced from $Pnma$ (space group 62) to $P2_1ma$ (space group 26). Although the system remains orthorhombic, such change in crystal symmetry would be associated with a change in the diffraction pattern and should therefore be identifiable, e.g. in an X-ray Laue diffraction experiment. Furthermore, tracking the change of symmetry as a function of field would allow to study whether the system stays in the lower symmetry phase as soon as it is entered or whether it alternates with field, e.g. with $Pnma$ in non-ME phases (II, III, V, VI) and $P2_1ma$ in ME phases (I, IV, VII). However, performing such an experiment at elevated fields is yet to be carried out and poses a future challenge. To investigate the possibility for changes in the crystal symmetry as a function of applied field, it would also be interesting to measure other components in the ME tensor such as e.g. $P_c$ for $H||c$. Previous \textit{ab initio} calculations show that the local single-ion anisotropy may be responsible for the canted magnetic structure that in turn enables the ME effect in LiNiPO$_4$ at low fields applied along $a$ \cite{yamauchi2010}. It would be interesting to perform such study for high magnetic fields along $c$ in order to understand the ME effect in phases IV and VII. Here one would also have to take into account the potential change in crystal symmetry.

\section{Conclusions}

The magnetic phase diagram of LiNiPO$_4$ was characterized by magnetization, electric polarization up to $56\,\mathrm{T}$ and neutron diffraction up to $42\,\mathrm{T}$ applied along the easy axis $c$. In addition to already established transitions, we discover phase transitions at $37.6$. $39.4$ and $54\,\mathrm{T}$. Furthermore, magnetic structure refinements using integrated neutron intensities of Bragg peaks observed in phase V ($20.9-37.6\,\mathrm{T}$) indicate a circular spiral structure with spins in the $(a,c)$-plane. The spiral is propagating along $b$ and has a period of 3 nuclear unit cells. Phase VI ($37.6-39.4\,\mathrm{T}$) is very similar to phase V but for an increased magnetic susceptibility and possibly a slightly longer period of the magnetic structure. In phase VII ($39.4-54\,\mathrm{T}$), yet another commensurate magnetic structure is established. This phase displays a quadratic magnetoelectric effect and the proposed spin structure is similar to those found in the other magnetoelectric phases I and IV.

A generalized version of the model describing the field-induced electric polarization in phases I and IV is developed. The magnetoelectric effect in LiNiPO$_4$ is clearly connected to phases where the magnetic unit cell is identical to the crystallographic unit cell and we speculate on the underlying physical mechanism. 


Further experimental work is required to investigate the magnetic structure in phase VIII ($>54\,\mathrm{T}$) as well as to search for evidence for structural distortions.

\section*{Acknowledgements}

This work was supported by the Danish Agency for Science and Higher Education under DANSCATT. Neutron experiments were performed at the BER II research reactor at the Helmholtz-Zentrum Berlin, Germany, and at the Materials and Life Science Experimental Facility at J-PARC, Japan (Proposal No. 2015A0122). Magnetization and electric polarization measurements were performed at the Institute for Solid State Physics, Japan. We are grateful to Yusuke Matsuda and Hiromu Suto for assisting with the NOBORU experiment and we thank David Vaknin for providing crystals for all experiments. E. F., R. T.-P. and N. B. C. reminisce numerous enjoyable discussions with Niels Hessel Andersen on issues in condensed matter physics, including the lithium orthophosphates.


\begin{thebibliography}{10}

\bibitem{eerenstein2006}
W.~Eerenstein, N.~D. Mathur, and J.~F. Scott.
\newblock Multiferroic and magnetoelectric materials.
\newblock {\em Nature} {\bf 442}, 759--765, (2006).

\bibitem{cheong2007}
S.-W. Cheong and M.~Mostovoy.
\newblock Multiferroics: a magnetic twist for ferroelectricity.
\newblock {\em Nature Materials} {\bf 6}, 13--20, (2007).

\bibitem{rivera2009}
J.-P. Rivera.
\newblock A short review of the magnetoelectric effect and related experimental
  techniques on single phase (multi-) ferroics.
\newblock {\em The European Physical Journal B} \textbf{71}, 299--313, (2009).

\bibitem{fusil2014}
S.~Fusil, V.~Garcia, A.~Barthélémy, and M.~Bibes.
\newblock Magnetoelectric devices for spintronics.
\newblock {\em Annu. Rev. Mater. Res.} {\bf 44}, 91–116, (2014).

\bibitem{katsura2005}
H.~Katsura, N.~Nagaosa, and A.~V. Balatsky.
\newblock Spin current and magnetoelectric effect in noncollinear magnets.
\newblock {\em Phys. Rev. Lett.} \textbf{95}, 057205, (2005).

\bibitem{sergienko2006}
I.~A. Sergienko and E.~Dagott.
\newblock Role of the Dzyaloshinskii-Moriya interaction in multiferroic
  perovskites.
\newblock {\em Phys. Rev. B} {\bf 73}, 094434, (2006).

\bibitem{kim2014}
J.~W. Kim, S.~Khim, S.~H. Chun, Y.~Jo, L.~Balicas, H.~T. Yi, S.-W. Cheong,
  N.~Harrison, C.~D. Batista, J.~H. Han, and K.~H. Kim.
\newblock Manifestation of magnetic quantum fluctuations in the dielectric
  properties of a multiferroic.
\newblock {\em Nat. Commun.} {\bf 5}, 4419, (2014).

\bibitem{mostovoy2006}
M.~Mostovoy.
\newblock Ferroelectricity in spiral magnets.
\newblock {\em Phys. Rev. Lett.} {\bf 96}, 067601, (2006).

\bibitem{kimura2007}
T.~Kimura.
\newblock Spiral magnets as magnetoelectrics.
\newblock {\em Annu. Rev. Mater. Res.} {\bf 37}, 387, (2007).

\bibitem{kimura2003}
T.~Kimura, T.~Goto, H.~Shintani, K.~Ishizaka, T.Arima, and Y.Tokura.
\newblock Magnetic control of ferroelectric polarization.
\newblock {\em Nature} {\bf 426}, 55--58, (2003).

\bibitem{goto2004}
T.~Goto, T.~Kimura, G.~Lawes, A.~P. Ramirez, and Y.Tokura.
\newblock Ferroelectricity and giant magnetocapacitance in perovskite
  rare-earth manganites.
\newblock {\em Phys. Rev. Lett.} {\bf 92}, 257201, (2004).

\bibitem{kenzelmann2005}
M.~Kenzelmann, A.~B. Harris, S.~Jonas, C.~Broholm, J.~Schefer, S.~B. Kim, C.~L.
  Zhang, S.-W. Cheong, O.~P. Vajk, and J.~W. Lynn.
\newblock Magnetic inversion symmetry breaking and ferroelectricity in
  {TbMnO$_3$}.
\newblock {\em Physical Review Letters} {\bf 95}, 087206, (2005).

\bibitem{hur2004}
N.~Hur, S.~Park, P.~A. Sharma, S.~Guha, and S.-W. Cheong.
\newblock Colossal magnetodielectric effects in {DyMn$_2$O$_5$}.
\newblock {\em Phys. Rev. Lett.} {\bf 93}, 107207, (2004).

\bibitem{blake2005}
G.~R. Blake, L.~C. Chapon, P.~G. Radaelli, S.~Park, N.~Hur, S.-W. Cheong, and
  J.~Rodríguez-Carvajal.
\newblock Spin structure and magnetic frustration in multiferroic
  {$R$Mn$_2$O$_5$ ($R$ = Tb,Ho,Dy)}.
\newblock {\em Phys. Rev. B} {\bf 71}, 214402, (2005).

\bibitem{park2007}
S.~Park, Y.~J. Choi, C.~L. Zhang, and S.-W. Cheong.
\newblock erroelectricity in an {$S = 1/2$} chain cuprate.
\newblock {\em Phys. Rev. Lett.} {\bf 98}, 057601, (2007).

\bibitem{schrettle2008}
F.~Schrettle, S.~Krohns, P.~Lunkenheimer, J.~Hemberger, N.~Büttgen, H.~A.~Krug
  von Nidda, V.~Prokofiev, and A.Loidl.
\newblock Switching the ferroelectric polarization in the {$S=1/2$} chain
  cuprate {LiCuVO$_4$} by external magnetic fields.
\newblock {\em Phys. Rev. B} {\bf 77}, 144101, (2008).

\bibitem{brockhouse1953}
B.~N. Brockhouse.
\newblock Antiferromagnetic structure in {Cr$_2$O$_3$}.
\newblock {\em J. Chem. Phys.} \textbf{21}, 961, (1953).

\bibitem{astrov1961}
D.~N. Astrov.
\newblock Magnetoelectric effect in chromium oxide.
\newblock {\em J. Exp. Theor. Phys.} {\bf 40}, 1035-1041, (1961).

\bibitem{mays1963}
J.~M. Mays.
\newblock Nuclear magnetic resonances and Mn-O-P-O-Mn superexchange linkages in
  paramagnetic and antiferromagnetic {LiMnPO$_4$}.
\newblock {\em Phys. Rev.} {\bf 131}, 38--53, (1963).

\bibitem{santoro1966}
R.~P. Santoro, D.~J. Segal, and R.~E. Newnham.
\newblock Magnetic properties of {LiCoPO$_4$} and {LiNiPO$_4$}.
\newblock {\em J. Phys. Chem. Solids} {\bf 27}, 1192--1193, (1966).

\bibitem{santoro1967}
R.~P. Santoro and R.~E. Newnham.
\newblock Antiferromagnetism in {LiFePO$_4$}.
\newblock {\em Acta Cryst.} {\bf 22}, 344--347, (1967).

\bibitem{mercier}
M.~Mercier.
\newblock {\em Étude de l'effet magnetoelectrique sur de composés de type
  olivine, perovskite et grenat}.
\newblock PhD thesis, Université de Grenoble, 1969.

\bibitem{jensen2009_2}
T.~B.~S. Jensen, N.~B. Christensen, M.~Kenzelmann, H.~M. Rønnow,
  C.~Niedermayer, N.~H. Andersen, K.~Lefmann, J.~Schefer, M.~Zimmermann, J.~Li,
  J.~L. Zarestky, and D.~Vaknin.
\newblock Field-induced magnetic phases and electric polarization in
  {LiNiPO$_4$}.
\newblock {\em Phys. Rev. B} {\bf 79}, 092412, (2009).

\bibitem{toftpetersen2017}
R.~Toft-Petersen, E.~Fogh, T.~Kihara, J.~Jensen, K.~Fritsch, J.~Lee, G.~E.
  Granroth, M.~B. Stone, D.~Vaknin, H.~Nojiri, and N.~B. Christensen.
\newblock Field-induced reentrant magnetoelectric phase in {LiNiPO$_4$}.
\newblock {\em Phys. Rev. B} {\bf 95}, 064421, (2017).

\bibitem{toftpetersen2015}
R.~Toft-Petersen, M.~Reehuis, T.~B.~S. Jensen, N.~H. Andersen~J. Li, M.~D. Le,
  M.~Laver, C.~Niedermayer, B.~Klemke, K.~Lefmann, and D.~Vaknin.
\newblock Anomalous magnetic structure and spin dynamics in magnetoelectric
  {LiFePO$_4$}.
\newblock {\em Phys. Rev. B} {\bf 92}, 024404, (2015).

\bibitem{abrahams1993}
I.~Abrahams and K.~S. Easson.
\newblock Structure of lithium nickel phosphate.
\newblock {\em Acta Crystallographica Section C} \textbf{49}, 925--926, (1993).

\bibitem{vaknin2004}
D.~Vaknin, J.~L. Zarestky, J.-P. Rivera, and H.~Schmid.
\newblock Commensurate-incommensurate magnetic phase transition in
  magnetoelectric single crystal {LiNiPO$_4$}.
\newblock {\em Phys. Rev. Lett.} {\bf 92}, 207201, 2004.

\bibitem{toftpetersen2011}
R.~Toft-Petersen, J.~Jensen, T.~B.~S. Jensen, N.~H. Andersen, N.~B.
  Christensen, C.~Niedermayer, M.~Kenzelmann, M.~Skoulatos, M.~D. Le,
  K.~Lefmann, S.~R. Hansen, J.~Li, J.~L. Zarestky, and D.~Vaknin.
\newblock High-field magnetic phase transitions and spin excitations in
  magnetoelectric {LiNiPO$_4$}.
\newblock {\em Phys. Rev. B} {\bf 84}, 054408, (2011).

\bibitem{peedu2019}
L.~Peedu, V.~Kocsis, D.~Szaller, J.~Viirok, U.~Nagel, T.~Rõõm, D.~G.~Farkas, S.~Bordács, D.~L.~Kamenskyi, U.~Zeitler, Y.~Tokunaga, Y.~Taguchi, Y.~Tokura, and I.~Kézsmárki.
\newblock Spin excitations of magnetoelectric {LiNiPO$_ 4$} in multiple magnetic phases.
\newblock {\em Phys. Rev. B} {\bf 100}, 024406, (2019).

\bibitem{khrustalyov2016_Ni}
V.~M. Khrustalyov, V.~M. Savytsky, and M.~F. Kharchenko.
\newblock Magnetoelectric effect in antiferromagnetic {LiNiPO$_4$} in pulsed
  magnetic fields.
\newblock {\em Low Temp. Phys.} {\bf 42}, 1126--1129, (2016).

\bibitem{takeyama2010}
S.~Takeyama and K.~Kindo.
\newblock International megagauss science laboratory atthe institute for solid
  state physics, University of Tokyo.
\newblock {\em J. Phys.: Conf. Ser.} \textbf{51}, 663, (2010).

\bibitem{zherlitsyn2012}
S.~Zherlitsyn, B.~Wustmann, T.~Herrmannsdörfer, and J.~Wosnitza.
\newblock Status of the pulsed-magnet-development program at the Dresden High
  Magnetic Field Laboratory.
\newblock {\em IEEE. Trans. Appl. Supercond.} {\bf 22}, 4300603, (2012).

\bibitem{nguyen2016}
D.~N. Nguyen, J.~Michel, and C.~H. Mielke.
\newblock Status and development of pulsed magnets at the NHMFL pulsed field
  facility.
\newblock {\em IEEE. Trans. Appl. Supercond.} {\bf 26}, 4300905, (2016).

\bibitem{yoshii2009}
S.~Yoshii, K.~Ohoyama, K.~Kurosawa, H.~Nojiri, M.~Matsuda, P.~Frings, F.~Duc,
  B.~Vignolle, G.~L. J.~A. Rikken, L.-P. Regnault, S.~Michimura, and F.~Iga.
\newblock Neutron diffraction study on the multiple magnetization plateaus in
  {TbB$_4$} under pulsed high magnetic field.
\newblock {\em Phys. Rev. Lett.} {\bf 103}, 077203, (2009).

\bibitem{nojiri2011}
H.~Nojiri, S.~Yoshii, M.~Yasui, K.~Okada, M.~Matsuda, J.-S. Jung, T.~Kimura,
  L.~Santodonato, G.~E. Granroth, K.~A. Ross, J.~P. Carlo, and B.~D. Gaulin.
\newblock Neutron laue diffraction study on the magnetic phase diagram of
  multiferroic {MnWO$_4$} under pulsed high magnetic fields.
\newblock {\em Phys. Rev. Lett.} {\bf 106}, 237202, (2011).

\bibitem{smeibidl2016}
P.~Smeibidl, M.~Bird, H.~Ehmler, I.~Dixon, J.~Heinrich, M.~Hoffmann,
  S.~Kempfer, S.~Bole, J.~Toth, O.~Prokhnenko, and B.~Lake.
\newblock First hybrid magnet for neutron scattering at Helmholtz-Zentrum
  Berlin.
\newblock {\em IEEE Trans. Appl. Supercond.} {\bf 26}, 4301606, (2016).

\bibitem{prokhnenko2015}
O.~Prokhnenko, W.-D. Stein, H.-J. Bleif, M.~Fromme, M.~Bartkowiak, and
  T.~Wilpert.
\newblock Time-of-flight extreme environment diffractometer at the
  Helmholtz-Zentrum Berlin.
\newblock {\em Rev. Sci. Instrum.} {\bf 86}, 033102, (2015).

\bibitem{prokhnenko2016}
O.~Prokhnenko, M.~Bartkowiak, W.-D. Stein, N.~Stuesser, H.-J. Bleif, M.~Fromme,
  K.~Prokes, P.~Smeibidl, M.~Bird, and B.~Lake.
\newblock {HFM-EXED} - the high field facility for neutron scattering at {HZB}.
\newblock {\em Proceedings of ICANS-XXI}, pages 278--285, (2016).

\bibitem{prokhnenko2017}
O.~Prokhnenko, P.~Smeibidl, W.~D. Stein, M.~Bartkowiak, and N.~Stuesser.
\newblock {HFM/EXED}: The high magnetic field facility for neutron scattering
  at ber ii.
\newblock {\em JLSRF} {\bf 3}, A115, (2017).

\bibitem{akaki2012}
M.~Akaki, H.~Iwamoto, T.~Kihara, M.~Tokunaga, and H.~Kuwahara.
\newblock Multiferroic properties of an åkermanite {Sr$_2$CoSi$_2$O$_7$}
  single crystal in high magnetic fields.
\newblock {\em Phys. Rev. B} {\bf 86}, 060413(R), (2012).

\bibitem{mitamura2007}
H.~Mitamura, S.~Mitsuda, S.~Kanetsuki, H.~A. Katori, T.~Sakakibara, and
  K.~Kindo.
\newblock Dielectric polarization measurements on the antiferromagnetic
  triangular lattice system {CuFeO$_2$} in pulsed high magnetic fields.
\newblock {\em J. Phys. Soc. Jpn.} {\bf 76}, 094709, (2007).

\bibitem{schultz1982}
A.~J. Schultz, R.~G. Teller, S.~W. Peterson, and J.~M. Williams.
\newblock Collection and analysis of single crystal time-of-flight neutron
  diffraction data.
\newblock {\em AIP Conference Proceedings} {\bf 89}, 35--41, (1982).

\bibitem{fogh2017}
E.~Fogh, R.~Toft-Petersen, E.~Ressouche, C.~Niedermayer, S.~L. Holm,
  M.~Bartkowiak, O.~Prokhnenko, S.~Sloth, F.~W. Isaksen, D.~Vaknin, and N.~B.
  Christensen.
\newblock Magnetic order, hysteresis, and phase coexistence in magnetoelectric
  {LiCoPO$_4$}.
\newblock {\em Phys. Rev. B} {\bf 96}, 104420, (2017).

\bibitem{kharchenko2010}
N.~F. Kharchenko, V.~M. Khrustalev, and V.~N. Savitskii.
\newblock Magnetic field induced spin reorientation in the strongly anisotropic
  antiferromagnetic crystal {LiCoPO$_4$}.
\newblock {\em Low Temp. Phys.} {\bf 36}, 558--564, (2010).

\bibitem{arnold2014}
O.~Arnold, J.~C. Bilheux, J.~M. Borreguero, A.~Buts, S.~I. Campbell~L. Chapon,
  M.~Doucet, N.~Draper, R.~Ferraz Leal, M.~A. Gigg, V.~E. Lynch,
  A.~Markvardsen, D.~J. Mikkelson, R.~L. Mikkelson, R.~Miller, K.~Palmen,
  P.~Parker, G.~Passos, T.~G. Perring, P.~F. Peterson, S.~Ren, M.~A. Reuter,
  A.~T. Savici, J.~W. Taylor, R.~J. Taylor, R.~Tolchenov, W.~Zhou, and
  J.~Zikovsky.
\newblock Mantid -- data analysis and visualization package for neutron
  scattering and $\mu$SR experiments.
\newblock {\em Nucl. Instrum. Methods Phys. Res.} {\bf 764}, 156 -- 166,
  (2014).

\bibitem{rodriguezcarvajal1993}
J.~Rodriguez-Carvajal.
\newblock Recent advances in magnetic structure determination by neutron powder
  diffraction.
\newblock {\em Phys. B} \textbf{192}, 55, (1993).

\bibitem{grimmerInternationalTablesD}
A.~S. Borovik-Romanov, H.~Grimmer, and M.~Kenzelmann.
\newblock Magnetic properties.
\newblock In A.~Authier, editor, {\em International Tables for Crystallography
  Volume D: Physical properties of crystals}, pages 139--145. Springer,
  Dordrecht, (2013).

\bibitem{grimmer1994}
H.~Grimmer.
\newblock The forms of tensors describing magnetic, electric and toroidal
  properties.
\newblock {\em Ferroelectrics} {\bf 161}, 181--189, (1994).

\bibitem{goodenough}
J.~Goodenough.
\newblock {\em Magnetism and the Chemical Bond}.
\newblock New York: Wiley, (1963).

\bibitem{yamauchi2010}
K.~Yamauchi and S.~Picozzi.
\newblock Magnetic anisotropy in liphosphates and origin of magnetoelectricity
  in {LiNiPO$_4$}.
\newblock {\em Phys. Rev. B} {\bf 81}, 024110, (2010).

\end{thebibliography}
\end{document}